\documentclass[aps,11pt,showkeys,nofootinbib]{revtex4}

\usepackage{amsmath,amssymb,amsfonts,amsthm}
\usepackage{amsbsy} 
\usepackage{epsfig}
\usepackage{latexsym}

\usepackage{hyperref}

\newcommand{\oarX}[1]{\href{http://arxiv.org/abs/#1}{{\ttfamily #1}}}
\newcommand{\arX}[1]{\href{http://arxiv.org/abs/#1}{{\ttfamily arXiv:#1}}}
\newcommand{\doin}[2]{\href{http://dx.doi.org/#1}{#2}}

\def\be{\begin{equation}}
\def\ee{\end{equation}}
\def\dd{{\rm d}}
\def\bes{\begin{eqnarray}}
\def\ees{\end{eqnarray}}

\DeclareMathOperator{\tr}{tr}

\newcommand{\C}{\mathbb{C}}
\newcommand{\ii}{\text{i}}
\newcommand{\dee}{\text{d}}

\renewcommand{\bar}{\overline}

\begin{document}

\title{Unimodular Pleba\'{n}ski gravity}
\author{Steffen Gielen}
\email{s.c.gielen@sheffield.ac.uk}
\author{Elliot Nash}
\email{enash4@sheffield.ac.uk}
\affiliation{School of Mathematics and Statistics, University of Sheffield, Hicks Building, Hounsfield Road, Sheffield S3 7RH, United Kingdom}
\date{\today}

\begin{abstract}
We present new action principles for unimodular gravity, defined in the chiral Pleba\'{n}ski formulation based on (complex) two-forms and a complex ${\rm SO}(3)$ connection. In these theories, just as in their analogues in the metric formulation, the cosmological constant does not take a prescribed value but is an integration constant whose value can differ between different (classical) solutions. We discuss some subtleties when identifying Lorentzian solutions in the generally complex theory, and show how these theories can be reduced to a ``pure connection'' form similar to Krasnov's pure connection formalism for general relativity.
\end{abstract}


\maketitle

\section{Introduction}

General relativity is most commonly thought of as a diffeomorphism-covariant theory of Lorentzian metrics whose curvature tensors satisfy the Einstein equations, but there exist a large number of known formulations in terms of different variables or different symmetry principles. Rather than working with the metric, lengths and angles can be measured with a tetrad (frame) field, and curvature can be encoded in an independent connection that may have torsion or non-metricity. The tetrad itself can be encoded into a set of two-forms, or integrated out completely. Many of these formalisms are reviewed in \cite{Peldan} and in particular in the recent textbook \cite{KrasnovBook}. The diffeomorphism symmetry can be extended to Weyl transformations by including an additional ``dilaton'' scalar field, or reduced to a set of volume-preserving diffeomorphisms by fixing a preferred volume form \cite{MoreorLess} (see also \cite{parentTH}). Introducing a preferred volume form leads to {\em unimodular} gravity, which differs in some important aspects from usual general relativity; most importantly, the cosmological constant appears as an integration constant rather than a fundamental parameter of Nature \cite{Einstein1919,AndersonFinkelstein,Unruh}. For a single classical solution, unimodular gravity is indistinguishable from general relativity, but there has been some debate on equivalence at the quantum level or implications for the so-called cosmological constant problem \cite{Percacci,Smolin,Jirousek,UGReview,Alvarez}.\footnote{Many arguments point towards equivalence at least of a perturbative quantisation of unimodular gravity and general relativity \cite{UGReview,OhtaEtAl}.}

In this short note, we show that the (chiral) Pleba\'{n}ski formulation of general relativity \cite{KrasnovBook,Plebanski,PlebanskiIntro} admits various straightforward extensions to unimodular gravity, mirroring the ideas formulated traditionally in the metric theory: one can either introduce a preferred volume form and reduce the symmetry of the theory from the full diffeomorphism group to volume-preserving diffeomorphisms; or one can follow a ``parametrised'' formalism {\em \`a la} Henneaux--Teitelboim \cite{HenneauxTeitelboim} in which the cosmological constant $\Lambda$ is promoted to a dynamical field, with a Lagrange multiplier enforcing its constancy. In this latter approach, first dicussed in \cite{SmolinUG}, the volume form is related to this new Lagrange multiplier.

Apart from intrinsic interest in defining yet another way of formulating the dynamics of general relativity, the actions presented here have various interesting applications. First of all, they allow for the construction of new ``pure connection'' theories analogous to those found by Krasnov \cite{KrasnovPureConn}. The unimodular pure connection actions are in some sense more immediately defined than the one proposed in \cite{KrasnovPureConn}, which requires picking an arbitrary volume form to define the Lagrangian; the unimodular setting provides a preferred volume form that can be used for this purpose. The more minimalistic viewpoint offered by unimodular gravity compared to standard general relativity -- the overall conformal factor present in general relativity carries no degrees of freedom, so might be seen as redundant \cite{MoreorLess,vanderBij:1981ym} -- carries over to these connection-based formulations. 

Given that the traditional Pleba\'{n}ski formulation is the basis for attempts to quantise gravity through spin foams \cite{Perez}, the new actions also present a route towards definition of spin foam models for unimodular gravity. In particular, the interpretation of the simplicity constraints of Pleba\'{n}ski gravity changes in the unimodular approach, where they involve the preferred volume form. It would be very interesting to study how this new viewpoint could be implemented at the discrete level of spin foams, leading potentially to new models which are not equivalent to the traditionally studied models.

\section{Chiral Pleba\'{n}ski gravity}

The (Lorentzian) chiral Pleba\'{n}ski action for gravity in four dimensions is given by \cite{KrasnovBook}\footnote{Lowercase indices $i,j$ run from 1 to 3 whereas uppercase indices $I,J$ will run from 0 to 3.}
\begin{equation}
    \label{Plebanski action}
    S [ A , \Sigma , M , \omega] = \frac{-\ii}{8\pi G} \int \Sigma_i \wedge F^i - \frac{1}{2} M_{ij} \Sigma^i \wedge \Sigma^j + \frac{1}{2} \omega \left( \tr M - \Lambda \right)\,,
\end{equation}
where $\Sigma^i$ is a 2-form valued in the complex Lie algebra $\mathfrak{so}(3,\C)$, $F^i$ are the components of the curvature of an $SO(3,\C)$ connection $A^i$, $M_{ij}$ is a symmetric matrix field and $\omega$ is a complex 4-form (to be identified, up to a constant, with the spacetime volume form). $\Lambda$ is a fixed parameter representing the cosmological constant, usually taken to be real; $G$ denotes Newton's constant.

The field equations of chiral Pleba\'{n}ski gravity are (indices are raised and lowered by the Kronecker delta $\delta^{ij}$)
\begin{equation}
F^i=M^{ij}\Sigma_j\,,\quad D_A \Sigma^i=0\,,\quad \Sigma^i\wedge\Sigma^j=\delta^{ij}\omega\,,\quad \tr M =\Lambda
\label{plebanskieom}
\end{equation}
where $D_A$ is the exterior covariant derivative (here in the adjoint representation) with respect to the connection $A$. The third equation says that the 2-forms $\Sigma^i$ encode a tetrad $E^I$ and hence a metric; for non-vanishing $\omega$ the general solution is given by
\begin{equation}
\Sigma^i = \ii E^0\wedge E^i - \frac{1}{2}{\epsilon^i}_{jk}E^j\wedge E^k
\label{tetrad}
\end{equation}
in terms of some $E^I$ such that $\omega=-2\ii\,E^0\wedge E^1\wedge E^2\wedge E^3$. This result is well-known in the literature \cite{selfdual2}; we sketch the proof in the Appendix for completeness.

The second equation of motion then says that $A$ is torsion-free; the first equation says that the curvature of $F$ is self-dual (i.e., a linear combination only of the $\Sigma^i$ and not of $\bar\Sigma^i$) which encodes the trace-free part of the Einstein equations $R_{\mu\nu}-\frac{1}{4}Rg_{\mu\nu}=0$; and finally, the last equation gives the missing Einstein equation, determining the Ricci scalar in terms of $\Lambda$ \cite{PlebanskiIntro}.

In order to obtain Lorentzian solutions from these general complex ones, one needs to impose reality conditions
\begin{equation}
\text{Re} \left( \Sigma^i \wedge \Sigma_i \right) = 0\,,\quad \Sigma^i \wedge \bar\Sigma^j  = 0\,.
\label{Reality conditions}
\end{equation}
Solutions to these split into four cases, whose associated tetrad $E^I$ (\ref{tetrad}) is real up to a phase factor $e^{\ii n\pi/4}$ where $n\in\{0,1,2,3\}$. The (Urbantke) metric $g_{\mu\nu}=\eta_{IJ}E^I_\mu E^J_\nu$ associated to $E^I$, where $\eta_{IJ}={\rm diag}(-1, +1,  +1,  +1)$, can have Lorentzian signature (either $(-+++)$ or $(+---)$) or be of the form $\ii$ times a Lorentzian metric. The same metric can be directly obtained from $\Sigma^i$ via the formula
\begin{equation}
g_\Sigma(\xi,\eta) \,\omega = -\frac{1}{3}\,\epsilon_{ijk}\,i_\xi\Sigma^i \wedge i_\eta\Sigma^j\wedge\Sigma^k
\label{Urbantke}
\end{equation}
where $\xi$ and $\eta$ are arbitrary vector fields and $i_\xi\Sigma^i:=\Sigma^i(\xi,\cdot)$ denotes the contraction of a 2-form with the vector field $\xi$; see \cite{KrasnovBook,PureConnQC} for more details. The fact that the metric can be reconstructed explicitly from the $\Sigma^i$ allows defining arbitrary matter couplings at least in principle, see, e.g., \cite{MatterCouple}.

\section{Unimodular extensions of Pleba\'{n}ski gravity}

In the metric formulation, the trace Einstein equation $R=4\Lambda$ is almost redundant; the Bianchi identities applied to the trace-free Einstein equations force the Ricci scalar $R$ to be a constant, so the trace part only determines the value of this constant. This observation is the basis of the unimodular formulation in which one does not impose the trace equation, and $\Lambda$ becomes an integration constant. It is  natural to ask whether a similar reasoning applies to the Pleba\'{n}ski formulation; can we recover the trace condition $\tr M={\rm const}$ from the other equations of motion?

If we assume the simplicity constraint $\Sigma^i\wedge\Sigma^j=\delta^{ij}\omega$, we know that $\Sigma^i$ are of the form (\ref{tetrad}). Using this, one can derive the identity
\begin{equation}
{{\Sigma^i}_\mu}^\rho\, {\Sigma^j}_{\rho\nu}=-\delta^{ij}g_{\mu\nu}+{\epsilon^{ij}}_k{\Sigma^k}_{\mu\nu}
\label{contraction}
\end{equation}
where $g_{\mu\nu}$ is the Urbantke metric, which is also used to raise the $\rho$ index on $\Sigma^i$. We then also have ${\Sigma^i}^{\mu\nu}\, {\Sigma^j}_{\mu\nu}=4\delta^{ij}$.

Now the Bianchi identity $D_A F^i=0$ and the first two equations of (\ref{plebanskieom}) imply that $\Sigma_j\wedge(D_AM^{ij})=0$ and hence
\begin{equation}
0={\Sigma^i}^{\mu\nu}\left({\Sigma_j}_{\mu\nu} D_\rho M^{ij}+{\Sigma_j}_{\nu\rho} D_\mu M^{ij}+{\Sigma_j}_{\rho\mu} D_\nu M^{ij}\right)=4\delta^{ij}D_\rho M_{ij}-2\delta^{ij}D_\rho M_{ij}
\end{equation}
where we write the exterior covariant derivative of $M^{ij}$ as $D_A M^{ij}=D_\rho M^{ij}\,{\rm d}x^\rho$ and used (\ref{contraction}). Contraction with $\delta^{ij}$ reduces the covariant derivative to a partial derivative, $\delta^{ij}D_\rho M_{ij}=\partial_\rho(\delta^{ij}M_{ij})$. Hence $\tr M$ must be a (complex) constant. Just as in metric general relativity, the trace Einstein equation (with undetermined value of the cosmological constant) can be derived from the other field equations together with the Bianchi identities and a non-degeneracy condition on the metric.

The simplest way of removing the constraint $\tr M=\Lambda$ as a dynamical equation is to replace the dynamical volume form $\omega$ by a fixed (background) volume form $\omega_0$. This turns (\ref{Plebanski action}) into the {\em unimodular Pleba\'{n}ski action}
\begin{equation}
    \label{Unimodular Plebanski action}
        S [ A , \Sigma , M ; \omega_0 ]  = \frac{-\ii}{8\pi G} \int \Sigma_i \wedge F^i - \frac{1}{2} M_{ij} \Sigma^i \wedge \Sigma^j + \frac{1}{2} \omega_0\,\tr M
\end{equation}
where the term $\propto \int \omega_0\,\Lambda$, which is now just an additive constant, was dropped.

(\ref{Unimodular Plebanski action}) leads to the equations of motion
\begin{equation}
F^i=M^{ij}\Sigma_j\,,\quad D_A \Sigma^i=0\,,\quad \Sigma^i\wedge\Sigma^j=\delta^{ij}\omega_0
\label{unimodulareom}
\end{equation}
which, by our previous discussion, correspond to the dynamical equations of unimodular gravity: the volume form $\omega_0=-2\ii\,E^0\wedge E^1\wedge E^2\wedge E^3$ is now a fixed background field, so that solutions to these equations correspond to solutions of complex general relativity (with some value of $\Lambda$) subject to the constraint of compatibility with $\omega_0$. $\Lambda$ is an integration constant, so that the theory contains all solutions for all possible values of $\Lambda$. If $\omega_0$ is chosen to be nowhere vanishing, only solutions with a nowhere vanishing volume form will be allowed. Instead of having full diffeomorphism symmetry, the theory is only invariant under (active) volume-preserving diffeomorphisms which would leave $\omega_0$ and hence (\ref{Unimodular Plebanski action}) invariant. All of this is familiar from discussions in the metric theory \cite{MoreorLess}.

In general, the integration constant $\Lambda$ can be complex, so the theory has more ``Lorentzian'' solutions than the usual Pleba\'{n}ski theory: recall that the reality conditions (\ref{Reality conditions}) imply that the Urbantke metric is of the form $g_{\mu\nu}=\sigma\tilde{g}_{\mu\nu}$, where $\sigma\in\{1,-1,\ii,-\ii\}$ and $\tilde{g}_{\mu\nu}$ is real of Lorentzian signature $(-+++)$. The trace Einstein equation $R_{\mu\nu}=\Lambda g_{\mu\nu}$ is equivalent to $\tilde{R}_{\mu\nu}=\sigma  \Lambda \tilde{g}_{\mu\nu}$ where $\tilde{R}_{\mu\nu}$ is the Ricci tensor of the Lorentzian metric $\tilde{g}_{\mu\nu}$, which is real. Thus, if $\Lambda$ is a fixed real parameter, there are no dynamical solutions with $\sigma=\pm\ii$ (but both $\sigma=\pm 1$ sectors are always present) \cite{PureConnQC}. In the unimodular approach $\Lambda$ cannot be fixed and so ``Lorentzian'' solutions for all four possible ``signatures'' $\sigma$ are present. One could add another reality condition $\text{Im} \tr M=0$ to exclude the imaginary solutions by hand.

We can obtain further actions by integrating out fields. Substituting the field equation $\Sigma^i=M_{ij}^{-1}F^j$ back into (\ref{Unimodular Plebanski action}) yields a unimodular first-order action in terms of only $A$, $M$ and $\omega_0$,
\begin{equation}
    \label{UFO action}
    S_{{\rm UFO}}[A, M; \omega_0] = \frac{-\ii}{16\pi G} \int M^{-1}_{ij} \, F^i \wedge F^j + \omega_0\,\tr M\,.
\end{equation}
This is the unimodular analogue of the first order chiral connection theory discussed in \cite{KrasnovBook,PureConnQC,herfray2016anisotropic}. The field equations $D_A \Sigma^i_F=0$ and $\Sigma^i_F\wedge\Sigma^j_F=\delta^{ij}\omega_0$, for $\Sigma^i_F:=(M^{-1})^{ij}F_j$, are equivalent to (\ref{unimodulareom}).

We can take another step and also remove the matrix field $M_{ij}$ from the theory, following the ideas of the pure connection approach \cite{KrasnovPureConn}. Define a matrix field $X^{ij}$ such that $F^i\wedge F^j=X^{ij}\,\omega_0$; we can then write the simplicity constraint equation as
\begin{equation}
(M^{-1})^{ik}F_k \wedge (M^{-1})^{jl}F_l = \delta^{ij}\,\omega_0\quad\Leftrightarrow\quad M^{-1}XM^{-1}={\rm Id}
\end{equation}
which has the solution(s) $M=\sqrt{X}$ (subject to the ambiguities in defining a matrix square root). Substituting this relation into (\ref{UFO action}) yields the {\em unimodular pure connection action}
\begin{equation}
    \label{UPC action}
    S_\mathrm{UPC} [ A ; \omega_0 ] = \frac{-\ii}{8\pi G} \int \omega_0 \tr \sqrt{X}\,.
\end{equation}
This yields a single field equation $D_A \left( ( X^{-1/2} )_{ij} F^j \right) = 0$, which may be rewritten as $D_A \Sigma_X ^i = 0$ defining $\Sigma_X ^i = ( X^{-1/2} )^{ij} F_j$ which by definition already satisfy $\Sigma_X ^i \wedge \Sigma_X ^j = \omega_0 \delta^{ij}$.
Then again, we have the same field equations as in (\ref{unimodulareom}).
The condition that before now was written $\dee \tr M = 0$ becomes $\dee \tr \sqrt{X} = 0$ in this formulation. Likewise, imaginary solutions could be excluded by demanding that $\tr \sqrt{X}$ should be real.

The relation of (\ref{UPC action}) to Krasnov's pure connection action \cite{KrasnovBook,KrasnovPureConn},
\begin{equation}
    S_{{\rm PC}}[A] = \frac{-\ii}{16\pi G\Lambda} \int \varepsilon_X\,\left(\tr \sqrt{\tilde{X}}\right)^2\,,
\end{equation}
where $\varepsilon_X$ is a fixed but arbitrary volume form such that $F^i\wedge F^j=\tilde{X}^{ij}\,\varepsilon_X$, can be understood as follows. The equation of motion derived from $S_{{\rm PC}}$ is
\begin{equation}
D_A\left(\tr\sqrt{\tilde{X}}(\tilde{X}^{-1/2})_{ij}F^j\right)=0\,;
\end{equation}
here the matrix $M$ of the (non-unimodular) chiral first order theory is recovered by defining $M_{\tilde{X}}:=\Lambda\,\sqrt{\tilde{X}}/\tr\sqrt{\tilde{X}}$. But the quantity $\tr\sqrt{\tilde{X}}$ can be set to any desired value by changing $\varepsilon_X$ (which was arbitrary) and hence $\tilde{X}^{ij}$. The choice for which $\tr\sqrt{\tilde{X}}=\Lambda$ is precisely the unimodular gauge we are in when defining (\ref{UPC action}). Conceptually, we might say that defining $S_{{\rm UPC}}$ is simpler than defining $S_{{\rm PC}}$ since $S_{{\rm UPC}}$ comes with a preferred volume form, so there is no need to choose an arbitrary $\varepsilon_X$. This more minimalistic viewpoint comes at the price of losing the full diffeomorphism symmetry of $S_{{\rm PC}}$. In particular, the theory defined by $S_{{\rm UPC}}$ is {\em not} in the class of diffeomorphism-invariant gauge theories discussed in \cite{GaugeApproach}.

\section{Parametrised Unimodular Pleba\'{n}ski gravity}

An interesting variant of unimodular gravity was proposed by Henneaux and Teitelboim \cite{HenneauxTeitelboim}. In this approach, one restores the full diffeomorphism symmetry of gravity by replacing the fixed volume form $\omega_0$ by an auxiliary dynamical field, which can be thought of as a vector density or a dual 3-form. The cosmological constant $\Lambda$ is promoted to a dynamical field, but required to be constant by an additional constraint. The Henneaux--Teitelboim action for (Lorentzian) ``parametrised unimodular'' gravity reads
\begin{equation}
    \label{HT action}
    S[g,\Lambda,\mathcal{T}]=\frac{1}{16\pi G}\int {\rm d}^4 x\left\{\sqrt{-g}\;(R-2\Lambda)+2\Lambda\partial_\mu\mathcal{T}^\mu\right\}
\end{equation}
where $\mathcal{T}^\mu$ is a vector density. Notice that the last term does not involve $\sqrt{-g}$. In addition to the Einstein equations, one finds equations of motion
\begin{equation}
\sqrt{-g}=\partial_\mu \mathcal{T}^\mu\,,\quad \partial_\mu\Lambda=0\,.
\end{equation}
The volume element ${\rm d}^4 x\;\sqrt{-g}$ is no longer fixed a priori but determined dynamically in terms of $\mathcal{T}$, and the action has full diffeomorphism symmetry. 

One can replace $\mathcal{T}$ by its dual 3-form, which we will call $T$; this viewpoint is more suitable when working with differential forms. We can then define an Henneaux--Teitelboim analogue of the Pleba\'{n}ski action (\ref{Plebanski action}), namely
\begin{equation}
    S[\Sigma, A, M, \omega, \Lambda, T] = \frac{-\ii}{8\pi G} \int \Sigma^i \wedge F_i - \frac{1}{2} M_{ij} \, \Sigma^i \wedge \Sigma^j + \frac{1}{2} \omega \left(\tr M - \Lambda \right) + \frac{1}{2}\dee T\,\Lambda\,.
\end{equation}
As in the metric case, $\Lambda$ is now promoted to a dynamical field but forced to be constant by the equation of motion $\dee\Lambda = 0$ coming from variation with respect to $T$. Variation with respect to $\Lambda$ leads to $\omega=\dee T$, so again the volume form is determined in terms of $T$. We see that $\omega$ is redundant and we could also work with the simpler  parametrised unimodular Pleba\'{n}ski action 
\begin{equation}
    \label{Parametrised Plebanski}
    S[\Sigma, A, M, T] = \frac{-\ii}{8\pi G} \int \Sigma^i \wedge F_i - \frac{1}{2} M_{ij} \, \Sigma^i \wedge \Sigma^j + \frac{1}{2} \dee T\,\tr M\,,
\end{equation}
which we could have also obtained by substituting $\omega_0=\dee T$ into (\ref{Unimodular Plebanski action}).\footnote{The same procedure applied to (\ref{HT action}) yields the {\em bona fide} unimodular gravity action $S[g,\mathcal{T}]=\frac{1}{16\pi G}\int{\rm d^4}x\left(\partial_\mu\mathcal{T}^\mu\right)R$.} This action was first discussed (without an imaginary prefactor) in \cite{SmolinUG}. All these actions are evidently diffeomorphism invariant and no longer depend on background fields.

The field equations arising from (\ref{Parametrised Plebanski}) are
\begin{equation}
F^i=M^{ij}\Sigma_j\,,\quad D_A \Sigma^i=0\,,\quad \Sigma^i\wedge\Sigma^j=\delta^{ij}\,\dd T\,,\quad \dd\tr M =0\,.
\end{equation}
In this formulation we get the condition $\dd\tr M = 0$ as a field equation, without requiring Bianchi identities or a non-degenerate volume form. For a non-degenerate volume form $\omega\propto\,\dd T$, we have seen already that the field equation $\dd\tr M=0$ would be redundant. Notice that the volume form is now required to be globally exact, so by Stokes' theorem the volume of any spacetime region could be evaluated by integrating the 3-form $T$ over its boundary.\footnote{In particular, the total volume of a compact spacetime without boundary would be zero. This curiosity might be more relevant in the Euclidean than in the Lorentzian setting, where compact manifolds are of limited interest.} This restriction on global quantities such as the spacetime volume is the analogue of constraints coming from a fixed volume form $\omega_0$  in the non-parametrised setting (see, e.g., \cite{MoreorLess}), and is related to the appearance of a new global degree of freedom in terms of $\Lambda$. For a spacetime manifold $M$ with two boundary hypersurfaces associated to times $t_0$ and $t_1$, the volume can be written as
\begin{equation}
\mathcal{V}_4(M) \propto \int_M \dd T = \int_{t=t_1} T - \int_{t=t_0} T\,.
\label{volumetime}
\end{equation}
The integral of $T$ over 3-dimensional hypersurfaces can be seen as defining a global ``volume'' time proportional to spacetime volume, as in \cite{HenneauxTeitelboim}.

The elimination of fields in the 3-form approach proceeds in the same way as in the preferred volume form approach, except we replace $\omega_0$ with $\dd T$ wherever it appears. We get actions
\begin{align}
    S_{{\rm PUFO}}[A, M, T] & = \frac{-\ii}{16\pi G} \int M^{-1}_{ij} \, F^i \wedge F^j + \dee T\,\tr M\,,\
    \label{Parametrised first order action}
\\  S_{{\rm PUPC}}[A, T] & = \frac{-\ii}{8\pi G} \int \dee T\,\tr \sqrt{X_T}\,,
\end{align}
where the matrix $X^{ij}_T$ is defined via $F^i \wedge F^j = \dee T\,X^{ij}_T$. This implies that for the variation,
\begin{align}
\dee T\,\delta X^{ij}_T = D_A\delta A^i\wedge F^j + F^i\wedge D_A\delta A^j - \dee\delta T\,X^{ij}_T\,.
\end{align}
The field equations arising from these actions are the same as those discussed below (\ref{UFO action}) (with $\omega_0$ replaced by $\dee T$) and (\ref{UPC action}), respectively, plus an additional equation $\dee\tr M=0$ or $\dee\left(\tr \sqrt{X_T}\right)=0$, which would anyway arise from the Bianchi identities for a non-degenerate $\dee T$.

\section{Minisuperspace model}

We can restrict these theories to a homogeneous isotropic Friedmann--Lema\^{i}tre--Robertson--Walker (FLRW) Universe,  as was done for pure connection cousins of Pleba\'{n}ski gravity in \cite{PureConnQC}.

Consider an FLRW universe with spatial slices of positive curvature  $k>0$ (the flat case can be obtained as a limit $k\rightarrow 0$ in all expressions below). One can fix a Cartan frame $e^{\bf a}$ satisfying $\dee e^{\bf a}+\sqrt{k}\,{\epsilon^{\bf a}}_{{\bf bc}}\, e^{\bf b} \wedge e^{\bf c}=0$. Then $\varepsilon_e = e^1\wedge e^2\wedge e^3$ defines a volume form and we have a ``fiducial'' (coordinate) volume $V=\int \varepsilon_e$. Variables of the model should be defined so as to not depend on the value of $V$, which is arbitrary.

We can work in formalisms where the 2-forms $\Sigma^i$ have been integrated out. Then a homogeneous isotropic ansatz for the connection $A^i$ and matrix $M_{ij}$ is given by
\begin{equation}
A^i = \left(V^{-1/3}\frac{\ii c}{3}+\sqrt{k}\right)e^i\,,\quad M_{ij}=\frac{c^2+9V^{2/3}k}{8\pi G\cdot 9p}\delta_{ij}\,.
\end{equation}

The parametrised unimodular action (\ref{Parametrised first order action}), which has the full diffeomorphism symmetry of general relativity, leads to a particularly transparent minisuperspace model. The other formulations we have discussed are dynamically equivalent, but require introduction of new variables or elimination of redundant ones. We may choose the 3-form $T$ to be given by
\begin{equation}
T = -(8\pi G)^2\cdot 2\ii \,V^{-1}\Pi\,\varepsilon_e\,.
\end{equation}
The free functions $c$, $p$ and $\Pi$ only depend on time $t$.

The curvature 2-form associated to this choice of $A^i$ is given by
\begin{equation}
F^i = \ii V^{-1/3}\frac{\dot{c}}{3} \dee t\wedge e^i - \frac{1}{2}{\epsilon^i}_{jk}e^j\wedge e^k\left(V^{-2/3}\frac{c^2}{9}+k\right)
\end{equation}
and the action (\ref{Parametrised first order action}) reduces to
\begin{equation}
S[c,p,\Pi]=\int \dd t\left(\dot{c}p+\dot\Pi\frac{c^2+9V^{2/3}k}{3p}\right)\,,
\end{equation}
which shows that the variable $p$ introduced in the ansatz for the matrix $M$ is just the canonical momentum for $c$. Moreover, the definition of $c$ is such that, if the dynamical variables $c$, $p$ and $\Pi$ are chosen to be real, the two-forms $\Sigma^i_F=(M^{-1})^{ij}F_j$ correspond to a real tetrad $E^I$ given by
\begin{equation}
E^i=V^{-1/3}\sqrt{8\pi G\,p}\,e^i\,,\quad E^0=\frac{3\sqrt{8\pi G\,p}}{c^2+9V^{2/3}k}\,\dd c=\frac{\sqrt{8\pi G}}{p^{3/2}}\,\dd\Pi
\end{equation}
using the equation of motion for $c$. The spacetime volume between two initial and final hypersurfaces becomes $\mathcal{V}_4=(8 \pi G)^2(\Pi(t_f)-\Pi(t_i))$ as discussed in the general case of (\ref{volumetime}).

In a Hamiltonian analysis, introducing a canonical momentum $\lambda$ for $\Pi$ gives a primary constraint
\begin{equation}
\lambda-\frac{c^2+9V^{2/3}k}{3p}\approx 0\,.
\end{equation}
This is the usual Hamiltonian constraint for the FLRW universe with dimensionless cosmological constant $\lambda=8\pi G\Lambda$ \cite{PureConnQC}. We can pass to an extended Hamiltonian action
\begin{equation}
S[c,p,\Pi,\lambda,N]=\int \dd t\left(\dot{c}p+\dot{\Pi}\lambda-N\left(\lambda-\frac{c^2+9V^{2/3}k}{3p}\right)\right)\,;
\label{newaction}
\end{equation}
this is the standard minisuperspace action for gravity with cosmological constant \cite{PureConnQC} (see also \cite{ImaginarySpace} for a more general viewpoint), with the cosmological constant promoted to a dynamical variable conjugate to $\Pi$. This addition of a new global degree of freedom associated to the cosmological constant is familiar from all unimodular extensions of general relativity.

We can immediately derive transition amplitudes or two-point functions encoding the quantisation of this model, using a path integral approach. Let us consider a two-point function for fixed boundary values of $c$ and $\lambda$. The structure of (\ref{newaction}) implies that this is given by
\begin{equation}
G(c_f,\lambda_f|c_i,\lambda_i) = \int \mathcal{D}\Pi\,\mathcal{D}\lambda\,\exp\left(\ii\int \dee t\;\dot\Pi\lambda\right)G_\lambda(c_f|c_i)=\delta(\lambda_f-\lambda_i)G_{\lambda_f}(c_f|c_i)
\end{equation}
where $G_\lambda(c_f|c_i)$ is the two-point function of the minisuperspace model with {\em fixed} $\lambda$. The result for the latter can be found in the literature \cite{Isichei,PureConnQC}, and depends on the type of two-point function one is interested in. For a two-point function that solves the Wheeler--DeWitt equation, one would integrate $N$ over the whole real line and find
\begin{equation}
G_{\lambda}(c_f|c_i) = \exp\left[\frac{\ii}{3\lambda}\left(\frac{c_f^3-c_i^3}{3}+9V^{2/3}k\,(c_f-c_i)\right)\right]\,.
\end{equation}
The discussion here is similar to the one given for unimodular gravity in \cite{Isichei}. Note that ``no boundary'' initial data (trying to set $p_i=0$ initially and using the constraint) would correspond to the two possible choices $c_i=\pm 3\ii\,V^{1/3}\sqrt{k}$ which (depending on the sign) lead to either exponential suppression or enhancement.

As a final remark, for a fixed $\lambda$ model the two-point function could be understood from the observation that the minisuperspace action could be reduced to a pure boundary term after inserting the solution for $p$ in terms of $c$ \cite{PureConnQC}. This reflected the fact that such a model has no independent degrees of freedom. In the unimodular setting we have a global degree of freedom given by $\lambda$, and no such reduction to a boundary term is possible.

\section{Conclusions}

We have shown how to incorporate the main ideas behind unimodular gravity -- a fixed volume form and a cosmological constant that appears as an integration constant, rather than a fundamental parameter -- into the (chiral) Pleba\'{n}ski formulation. Our constructions mirror closely what is done in the metric approach; one can either introduce a background field $\omega_0$ representing the volume form, or one can promote $\Lambda$ to a field constrained to be constant by a new Lagrange multiplier-like field which then also determines the volume form. Classically, these theories are equivalent to the usual theories based on the chiral Pleba\'{n}ski formulation, including a first order chiral connection theory obtained from integrating out the 2-form $\Sigma^i$, and the pure connection action studied by Krasnov. There is an important difference when trying to restrict to Lorentzian solutions via reality conditions; the usual reality conditions now allow for solutions corresponding to purely imaginary metrics, which could be excluded by hand by adding further constraints. 

All of the actions discussed above have a Euclidean signature counterpart, leading to tetrads encoding a positive-definite (or negative-definite) metric tensor.
To construct these actions, we simply require that all of the fields be real valued, and we omit the factor of $-\ii$ present in the constant prefactors.\footnote{This factor is absent in the discussion of \cite{SmolinUG} which might hence implicitly refer to Euclidean signature.}
For example when working with homogeneous spacetimes, which have a well-defined notion of time, it may be easiest to first work in the Euclidean regime where computations are simpler, and use a Wick rotation (or a choice of complex boundary conditions) to transform the results into the Lorentzian regime \cite{PureConnQC}.

While equivalence of the classical unimodular and standard formulations of general relativity is more or less straightforwardly understood, whether there is equivalence at the quantum level is more of an open question \cite{UGReview}. Since the Pleba\'{n}ski formulation and its descendants are the starting point for the construction of spin foam models for general relativity, one interesting question is whether one could now construct spin foam models for unimodular gravity. These might suggest different implementations of simplicity constraints, which involve the preferred volume form in the setting illustrated here. On a more foundational level, one key property of unimodular formulations is that they provide a globally valid time coordinate related to the total spacetime volume, and a Schr\"odinger-type interpretation of quantisation becomes available \cite{UnruhWald}. Quantisation of such models, whether through spin foams or otherwise, could hence be studied in terms of (potentially unitary) evolution in unimodular time.

There is also the question of how to incorporate matter into these formulations.
In the formulations in which the 2-forms $\Sigma^i$ are dynamical variables, for instance (\ref{Unimodular Plebanski action}) and (\ref{Parametrised Plebanski}), one may construct a metric tensor via the Urbantke formula (\ref{Urbantke}).
One may then use this metric to couple scalar and tensor fields in the usual way, as in \cite{MatterCouple}. The coupling of fermions was discussed in \cite{selfdual2}.
In the cases where $\Sigma^i$ are not independent variables, as in (\ref{UFO action}) or (\ref{UPC action}) for example, one may reconstruct $\Sigma^i$ via $\Sigma^i = (M^{-1})^{ij} F_j$ or $\Sigma^i = (X^{-1/2})^{ij} F_j$ respectively; one can then define the Urbantke metric and follow the same approach. While consistent coupling of matter fields can be achieved for all actions we consider, the complexity of the expression for the Urbantke metric increases as we descend the hierarchy from Pleba{\'n}ski-like to first-order and ultimately pure connection formulations. All this is very similar to the usual Pleba\'{n}ski formulation.

\acknowledgments
This work was supported by the Royal Society through the University Research Fellowship Renewal URF$\backslash$R$\backslash$221005 (SG). We are grateful to a referee for our previous paper \cite{PureConnQC} whose questions motivated us to look into the question of unimodular Pleba\'{n}ski formulations.

\appendix
\section{Solving the simplicity constraints}

This summary follows \cite{KrasnovBook,selfdual2}. Assuming $\Sigma^i\wedge\Sigma^j=\delta^{ij}\omega$ with non-zero $\omega$, construct two 2-forms
\begin{equation}
\Sigma^\pm = \Sigma^1 \pm \ii \Sigma^2\,.
\end{equation}
These are both simple, $\Sigma^+\wedge\Sigma^+=\Sigma^-\wedge\Sigma^-=0$, and so can be written as wedge products of two one-forms
\begin{equation}
\Sigma^+ = \theta^0\wedge\theta^1\,,\quad \Sigma^-=w^2\wedge w^3\,.
\end{equation}
We then also have $\Sigma^+\wedge\Sigma^-=\theta^0\wedge\theta^1\wedge w^2\wedge w^3 = 2\omega$. From $\Sigma^3\wedge\Sigma^+=\Sigma^3\wedge\Sigma^-=0$ we then get
\begin{equation}
\Sigma^3=\frac{1}{2}\left(O_{11}\,\theta^0\wedge w^3+O_{12}\,\theta^0\wedge w^2+O_{21}\,\theta^1\wedge w^3+O_{22}\,\theta^1\wedge w^2\right)
\end{equation}
with undetermined coefficients $O_{\alpha\beta}$. The last condition $\Sigma^3\wedge\Sigma^3=\omega$ tells us $O_{11}O_{22}-O_{12}O_{21}=1$ so these coefficients form a matrix $O\in{\rm SL}(2,\mathbb{C})$, which can be used to define new one-forms
\begin{equation}
\begin{pmatrix}\theta^3 \cr \theta^2\end{pmatrix}=O\begin{pmatrix}w^3 \cr w^2\end{pmatrix}\,,\quad \theta^2\wedge\theta^3= w^2\wedge w^3=\Sigma^-\,.
\end{equation}
In terms of this new basis, we find $\Sigma^3 = \frac{1}{2}\left(\theta^0\wedge \theta^3+\theta^1\wedge \theta^2\right)$. These one-forms define a tetrad via, e.g.,
\begin{equation}
\theta^0=-E^1-\ii\,E^2\,,\quad \theta^1=\ii(E^0-E^3)\,,\quad \theta^2=E^0+E^3\,,\quad\theta^3=\ii\,E^1+E^2
\end{equation}
such that (\ref{tetrad}) holds. By self-consistency we then also have $\omega=-2\ii\,E^0\wedge E^1\wedge E^2\wedge E^3$.

\end{document}